\documentclass[twocolumn,english]{revtex4}
\usepackage[T1]{fontenc}
\usepackage[utf8]{inputenc}
\usepackage{color}
\usepackage{amsmath}
\usepackage{amssymb}
\usepackage{graphicx}
\usepackage{esint}

\makeatletter

\newcommand{\noun}[1]{\textsc{#1}}

\@ifundefined{textcolor}{}
{%
 \definecolor{BLACK}{gray}{0}
 \definecolor{WHITE}{gray}{1}
 \definecolor{RED}{rgb}{1,0,0}
 \definecolor{GREEN}{rgb}{0,1,0}
 \definecolor{BLUE}{rgb}{0,0,1}
 \definecolor{CYAN}{cmyk}{1,0,0,0}
 \definecolor{MAGENTA}{cmyk}{0,1,0,0}
 \definecolor{YELLOW}{cmyk}{0,0,1,0}
}

\usepackage{babel}

\usepackage{babel}

\makeatother

\usepackage{babel}
\begin{document}

\title{Time-dependent phase shift of a retrieved pulse in off-resonant EIT-based
light storage}

\author{M.-A. Maynard, R. Bouchez, J. Lugani, F. Bretenaker, F. Goldfarb
and E. Brion }

\address{Laboratoire Aim\'{e} Cotton, CNRS, Universit\'{e} Paris Sud, ENS Cachan,
91405 Orsay, France.}
\begin{abstract}
We report measurements of the time-dependent phases of the leak and
retrieved pulses obtained in EIT storage experiments with metastable
helium vapor at room temperature. In particular, we investigate the
influence of the optical detuning at two-photon resonance, and provide
numerical simulations of the full dynamical Maxwell-Bloch equations,
which allow us to account for the experimental results.
\end{abstract}


\maketitle

\section{Introduction}

Because they do not interact with each other and can be guided via
optical fibers over long distances with relatively low losses, photons
appear as ideal information carriers and are therefore put forward
as the \textquotedbl{}flying qubits\textquotedbl{} in most of quantum
communication protocols. The design of memories able to reliably store
and retrieve photonic states is, however, still an open problem. The
most commonly studied protocol, considered to implement such a quantum
memory, is electromagnetically induced transparency (EIT) \cite{BIH91}.
This protocol was implemented in various systems such as cold atoms,
gas cells, or doped crystals \cite{LDB01,PFM01,HLL10}. Although the
Doppler broadening might seem to lead to strong limitations, EIT-based
light storage in warm alkali vapors gives good results and is still
a subject of active investigation \cite{NWX12}. In the last years,
some experiments were also performed in a Raman configuration, using
pulses which are highly detuned from the optical resonances in gas
cells \cite{RNL10,RML11,RNJ12}.

The EIT-based storage protocol in a $\Lambda$ atomic system relies
on the long-lived Raman coherence between the two ground states which
are optically coupled to the excited level. When a strong coupling
beam is applied along one of the two transitions, a narrow transparency
window limited by the Raman coherence decay rate is opened along the
other leg of the system. Because of the slow-light effect associated
with such a dramatic change of the medium absorption properties, a
weak probe pulse on the second transition is compressed while propagating
through the medium. When this pulse has fully entered the atomic medium,
it can be mapped onto the Raman coherences which are excited by the
two-photon process by suddenly switching off the coupling beam. It
can be safely stored during times smaller than the lifetime of Raman
coherence. Finally, the signal pulse can be simply retrieved by switching
on the coupling beam again. In the Raman configuration, the coupling
and probe pulses are optically far off-resonance but still fulfill
the two-photon transition condition. The advantage is a large bandwidth,
that allows to work with data rates higher than in the usual EIT regime
\cite{RNL10}.

Atoms at room temperature in a gas cell are particularly attractive
for light storage because of the simplicity of their implementation.
The effects of the significant Doppler broadening can be minimized
using co-propagating coupling and probe beams, so that the two-photon
resonance condition can be verified for all velocity classes: all
the atoms can thus participate to the EIT phenomenon as soon as they
are pumped in the probed level. As a consequence, handy simple gas
cells have turned out to be attractive for slow or even stopped light
experiments \cite{NWX12}. In a previous work \cite{MLM14}, we have
reported on an added phase shift recorded for EIT-based light storage
experiments carried out in a helium gas at room temperature when the
coupling beam is detuned from the center of the Doppler line. The
simple model that we have derived could not satisfactorily account
for our observations that were recorded for intermediate detunings,
e.g. close to the Doppler broadening of the transition. In the present
paper, we come back to this problem and provide new experimental results,
\emph{i.e.} time-dependent measurements of the retrieved signal phase
shift, as well as numerical results obtained through the simulation
of the full system of Maxwell-Bloch equations. The behaviour of these
phase shifts with the coupling detuning seems satisfactorily accounted
for by our simulations. We also perform numerical calculations in
the Raman regime.

The paper is organized as follows. In Section \ref{SecII} we present
the system and setup and describe how to measure the time-dependent
phase shift of the retrieved pulse with respect to the coupling beam.
We also briefly recall the system of Maxwell-Bloch equations which
governs our system and describe their numerical integration. In Section
\ref{SecIII}, we provide our experimental and numerical results and
show that they qualitatively agree. We also apply our simulations
to the far off-resonant Raman case. Finally, we conclude in Section
\ref{SecIV} and give possible perspectives of our work.

\section{Experimental setup and numerical simulations\label{SecII}}

\subsection{EIT storage experimental setup}

The atoms preferably used for EIT storage experiments are alkali atoms,
mainly rubidium and sometimes sodium or caesium. We choose here to
work with metastable $^{4}$He atoms, which have the advantage of
a very simple structure without hyperfine levels: transitions are
thus far enough one from another to investigate the effect of detunings
of the coupling and probe beams on light storage and retrieval. 

\begin{figure}
\begin{centering}
\includegraphics[width=6cm]{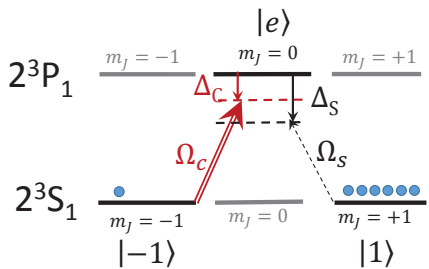}
\par\end{centering}

\protect\caption{\textcolor{black}{Atomic structure scheme for the D1 transition of
metastable helium. The relevant states which constitute the three-level
$\Lambda$ system are shown in black. $\Delta_{c}$ and $\Delta_{s}$
are the optical detunings, and $\Omega_{c}$ and $\Omega_{s}$ the
Rabi frequencies of the coupling and signal beams, respectively. The
two photon resonance is achieved for $\Delta_{c}=\Delta_{s}$.}}
\label{Atomic struture}
\end{figure}

In our setup represented in Fig. \ref{Experimental scheme}, a $6$-cm-long
cell is filled up with $1$ Torr of helium atoms which are continuously
excited to their metastable state $2^{3}S_{1}$ by a radio-frequency
(rf) discharge at 27 MHz. Each of the metastable ground states $\left|2^{3}S_{1},m_{J}=0,\pm1\right\rangle $
is hence fed with the same rate, denoted by $\frac{\Lambda}{3}$.
The cell is isolated from magnetic field gradients by a three-layer
$\text{\ensuremath{\mu}}$-metal shield to avoid spurious dephasing
effects on the different Zeeman components. A strong circularly-polarized
field, called the control beam, propagates along the quantization
axis $z$. Its power is set at $18$ mW for a beam diameter of $3$
mm. As shown in Fig. \ref{Atomic struture}, the coupling field drives
the transitions $\left|2^{3}S_{1},m_{J}=-1\right\rangle \leftrightarrow\left|2^{3}P_{1},m_{J}=0\right\rangle $
and $\left|2^{3}S_{1},m_{J}=0\right\rangle \leftrightarrow\left|2^{3}P_{1},m_{J}=1\right\rangle $.
Owing to the spontaneous transitions $\left|2^{3}P_{1},m_{J}=0\right\rangle \rightarrow\left|2^{3}S_{1},m_{J}=\pm1\right\rangle $
and $\left|2^{3}P_{1},m_{J}=1\right\rangle \rightarrow\left|2^{3}S_{1},m_{J}=0,1\right\rangle $,
the atoms end up in the state $\left|1\right\rangle \equiv\left|2^{3}S_{1},m_{J}=1\right\rangle $
within a few pumping cycles after the coupling beam has been switched
on. As the atoms are at room temperature, the Doppler broadening in
the cell is $W_{D}/2\pi\approx1\,\mbox{GHz}$. We denote by $\Delta_{c}\equiv\omega_{c}-\omega_{0}$
the detuning of the coupling frequency $\omega_{c}$ with respect
to the natural frequency $\omega_{0}/(2\pi)\approx2.8\times10^{14}\,\mbox{Hz}$
of the transition $2^{3}S_{1}\leftrightarrow2^{3}P_{1}$, at the center
of the Doppler line.

Once optical pumping is achieved, a weak signal pulse is sent through
the atomic medium along the $z$ axis. Its polarization is circular
and orthogonal to that of the coupling beam: the signal therefore
couples the state $\left|1\right\rangle $ to $\left|e\right\rangle \equiv\left|2^{3}P_{1},m_{J}=0\right\rangle $
and we denote by $\Delta_{s}\equiv\omega_{s}-\omega_{0}$ the detuning
of the signal frequency $\omega_{s}$ from the center of the Doppler
profile. Both signal and coupling beams are derived from the same
laser diode, and their frequencies and amplitudes are controlled by
two acousto-optic modulators. Due to the efficiency of optical pumping
through the coupling beam, we assume that the state $\left|2^{3}S_{1},m_{J}=0\right\rangle $
remains essentially unpopulated during the whole process and we accordingly
neglect the driving of the transition $\left|2^{3}S_{1},m_{J}=0\right\rangle \leftrightarrow\left|2^{3}P_{1},m_{J}=-1\right\rangle $
by the signal field. Submitted to the coupling and signal fields,
the atoms therefore essentially evolve in the three-level $\Lambda$
system $\left\{ \left|-1\right\rangle \equiv\left|2^{3}P_{1},m_{J}=-1\right\rangle ,\left|e\right\rangle ,\left|1\right\rangle \right\} $
(see Fig. \ref{Atomic struture}) as long as the detunings $\Delta_{c,s}=\omega_{c,s}-\omega_{0}$,
of the coupling and signal fields respectively, are small enough to
avoid exciting neighbouring transitions. Thanks to the absence of
hyperfine structure, the range of allowed values for $\Delta_{c,s}$
is, however, much larger than in alkali vapor experiments: indeed,
on the positive detuning side the nearest state ($^{3}P_{0}$) is
$30$ GHz away from optical resonance $\Delta_{c,s}=0$, while, on
the negative detuning side, the nearest state ($^{3}P_{2}$) is $2.29$
GHz away. 

\noindent \begin{center}
\begin{figure}
\begin{centering}
\includegraphics[width=8cm]{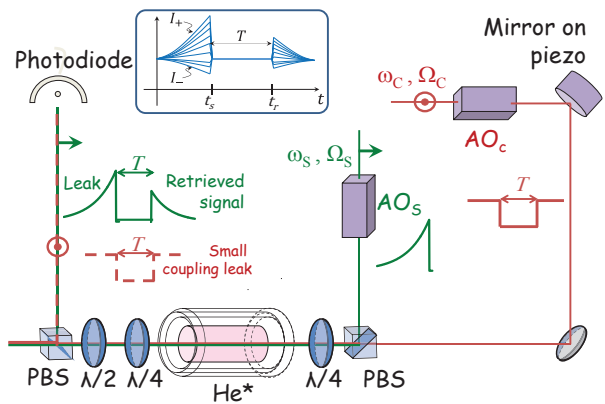}
\par\end{centering}

\protect\caption{\textcolor{black}{Experimental setup for EIT storage in metastable
helium. The coupling and signal beams are derived from the same laser
diode. They are initially linearly and orthogonally polarized, of
optical frequencies $\omega_{c}$ and $\omega_{s}$ and Rabi frequencies
$\Omega_{c}$ and $\Omega_{s}$, respectively. Acousto-optic modulators
are used to control the frequencies and amplitudes of the beams. Polarizing
Beam Splitters (PBS) allow to separate or recombine the beams. Circular
orthogonal polarizations are obtained by a quarter wave-plate. The
cell is contained into a $\text{\ensuremath{\mu}}$-metal shielding.
After the cell, polarization optics select mainly the probe beam with
some remaining coupling beam.The phase of the coupling beam is scanned
thanks to a mirror on a piezo-electric tranducer.}}

\label{Experimental scheme}
\end{figure}

\par\end{center}

Under EIT conditions, the coupling beam opens a transparency window
for the weak signal beam, which can therefore propagate without absorption
through the medium if its spectrum is not too wide \cite{FIM05}.
In the experimental results we present hereafter, we used a signal
pulse, which consists of a smoothly increasing exponential followed
by an abruptly decaying exponential of respective characteristic times
$2\,\mu\mbox{s}$ and $150$ ns. Its maximum power is about $170\,\mu\mbox{W}$
and the beam diameter is about $3$ mm. Different dissipative mechanisms
influence the width of the EIT window besides spontaneous emission,
such as collisions and transit of the atoms in and out of the beams.
These phenomena result in the decay of atomic coherences at the rates
$\gamma_{-11}/2\pi\approx14\,$kHz for the Raman coherence $\sigma_{-11}$
and $\gamma_{e1}/2\pi\approx22.8\mbox{ MHz}$ for the optical coherence
$\sigma_{e1}$. We have shown previously that velocity changing collisions
redistribute the pumping of the atoms over an effective width sligthly
smaller than the Doppler linewidth \cite{GGG09}. In our conditions,
with a coupling power of 18 mW, this effective width is experimentally
estimated to be $\Gamma_{D}/2\pi\approx0.8\,$GHz. Due to power broadening,
the width of the transparency window is then of the order of $500$
kHz.

\begin{figure}
\begin{centering}
\includegraphics[width=7.2cm]{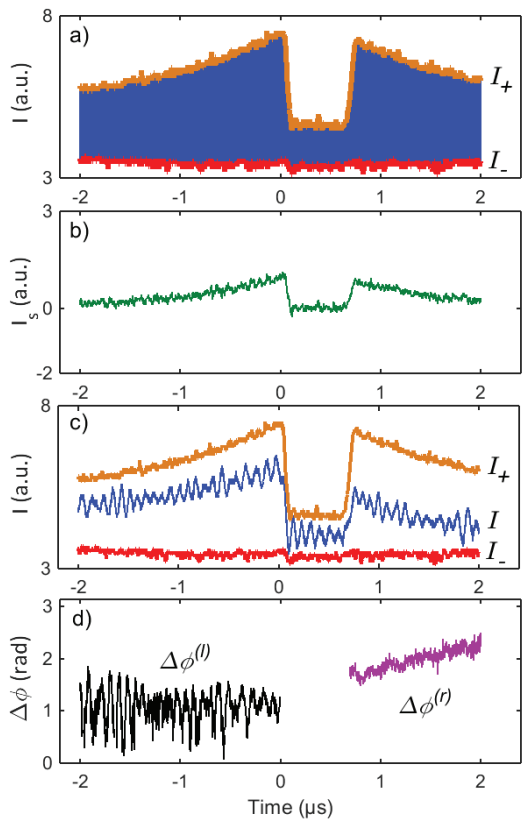}
\par\end{centering}

\protect\caption{(Color online) Experimental plots showing the data processing steps.
The coupling beam is switched off between times \textcolor{black}{$t=0$
and $t=0.6\,\mu$s. The optical detunings} are set at $\Delta_{c,s}=1$
GHz. The same vertical scale is used in graphs a, b and c. a) Accumulated
plot of the interference signal $I(t)$ between the probe and coupling
beams (blue). Upper and lower envelopes $I_{\pm}\left(t\right)$ are
shown in orange and red, respectively. b) Signal intensity $I_{s}\left(t\right)$
at the exit of the cell, deduced from the previous accumulated plot
a) and from a measurement of the coupling intensity $I_{c}$. c) Interference
signal $I(t)$ between the probe and coupling beams (blue). It is
contained between the upper and lower envelopes $I_{\pm}\left(t\right)$,
shown in orange and red, respectively. One can note the presence of
\textcolor{black}{spurious oscillations generated by the acousto-optic
modulators. d) Relative phases $\Delta\phi^{(l)}(t)$ (black) and
$\Delta\phi^{(r)}(t)$ (purple) between the signal and coupling beams
on the writing and retrieval parts, respectively. Notice that $\Delta\phi^{(l)}(t)$
is indeed constant.}}

\label{Data processing}
\end{figure}

The highly dispersive character of the medium under EIT conditions
can be used furthermore to store and retrieve a weak signal pulse:
due to EIT dispersion, the signal pulse indeed travels with a reduced
group velocity and is temporally contracted. Once the pulse has entered
the cell, the coupling beam can be switched off: information about
the signal pulse is then stored into the Raman coherences. After a
storage time $T$, the control beam is switched on again, which releases
the signal from the atomic coherence and ensures EIT absorption-free
propagation. Note that in our experimental setup, the optical depth
is only about 3.5 and the pulse can thus not be fully compressed in
the cell. Due to the finite width of the transparency window and the
finite length of the cell, a part, typically $10$ \% of the incoming
signal energy, leaks out before the coupling beam is turned off and
the storage period begins \cite{NWX12}.\noun{ }A typical experimental
record is given in Fig. \ref{Data processing}b. The first part of
the detected signal is the leak and after a storage time $T\approx0.6\,\mu$s,
once the coupling beam is turned on again, the retrieved signal is
clearly visible.

\subsection{Phase measurement setup}

In \cite{MLM14}, we investigated the relative phase $\Delta\phi$
of the signal with respect to the coupling beam, and showed the existence
of an optical detuning dependent extra phase shift $\varphi_{EIT}$
between the incident and retrieved pulse. This quantity can be measured
through mixing the signal emerging from the cell with a small fraction
of the control beam, via polarization optics. The resulting intensity
thus takes the form

\begin{equation}
I=I_{c}+I_{s}+\alpha\sqrt{I_{c}I_{s}\left(t\right)}\cos\left[\Delta\phi\right].\label{Intensity}
\end{equation}

In Eq. (\ref{Intensity}), the contrast factor $\alpha$, which ideally
equals $2$, accounts for non-perfect alignment of the beams and is
measured for each set of data. $I_{c}$ denotes the intensity of the
small fraction of the coupling field which is mixed and interferes
with the signal field. It takes the same constant value $I_{c}$ during
the writing and retrieval periods, while it vanishes during the storage
time. The value $I_{c}$ is measured in the absence of the signal
(one assumes that the introduction of the signal pulse does not substantially
affect the measurement of $I_{c}$). The phase of the coupling beam
is varied via a piezoelectric actuator from one experimental run to
another: the scan is slow enough so that it is assumed to be constant
during both the writing or retrieval steps. $I_{s}\left(t\right)$
is the time-dependent intensity of the signal beam emerging from the
cell. We denote by $I_{s}^{(l)}(t)$ and $I_{s}^{(r)}(t)$ the intensities
of the leak and retrieved pulses, respectively. Accordingly, we introduce
$\Delta\phi^{(l)}(t)$ and $\Delta\phi^{(r)}(t)$, the relative phases
of the leak and retrieved pulses, with respect to the coupling beam. 

To obtain the extra phase shift $\varphi_{EIT}$ between the incident
and retrieved pulses, we measure the relative phases $\Delta\phi^{(l)}$,
$\Delta\phi^{(r)}$ by homodyne detection. Repeating the same writing-storage-retrieval
sequence for many different postions of the piezoelectric actuator,
we obtain an accumulated plot whose upper/lower envelopes correspond
respectively to $I_{\pm}\left(t\right)=I_{c}+I_{s}\left(t\right)\pm\alpha\sqrt{I_{c}I_{s}\left(t\right)}$
(see Fig. \ref{Data processing}a). Given the previously measured
value of $I_{c}$, one can infer $I_{s}\left(t\right)$ from $\left(I_{+}+I_{-}\right)$
and $\alpha$ from $\left(I_{+}-I_{-}\right)$ (see Fig. \ref{Data processing}b).
For a given position of the piezo-actuator, one can then obtain $\Delta\phi^{(l)}(t)$
and $\Delta\phi^{(r)}(t)$ through fitting the experimental record
with Eq. (\ref{Intensity}) at each time $t$ (see Figs. \ref{Data processing}c,
d). It was verified both experimentally and numerically that the phase
of the leak is constant ($\Delta\phi^{(l)}(t)\equiv\Delta\phi^{(l)}$)
and can therefore be taken as a reference for the time-dependent relative
phase of the retrieved pulse $\Delta\phi^{(r)}(t)$. This time independence
of the leak phase is ensured by the fact that its spectral content
is much narrower than the EIT bandwidth. This is obtained thanks to
the shape of the signal pulse: a slow exponential increase, followed
by a sharp decrease. The part of the pulse which contains only low
frequencies enters first and gives form to a leak, whose phase is
constant. The ``extra phase shift'' is then measured as $\left[\Delta\phi^{(r)}(t)-\Delta\phi^{(l)}\right]\equiv\varphi_{EIT}(t)$.
Let us stress that in \cite{MLM14}, we assumed that the relative
phases $\Delta\phi^{\left(l,r\right)}$ of the leak and retrieved
pulses were time-independent: therefore, we directly extracted effective
``averaged'' values for $\alpha$ and $\varphi_{EIT}$ by performing
a two-parameter fit of the data with Eq. (\ref{Intensity}). Here,
by contrast, we measure the time-dependence of the phases and provide
experimental plots for $\varphi_{EIT}(t)$, without any assumption
on its behaviour.

\subsection{Numerical simulation principles}

For numerical simulations, we described the system in the one-dimensional
approximation. On the dimensions of the atomic sample, the coupling
and probe transverse profiles are assumed to remain constant. These
fields can therefore be cast under the form
\begin{eqnarray*}
\mathbf{E}_{c,s}\left(z,t\right) & = & \Re\left[\mathcal{E}_{c,s}\left(z,t\right)\mathbf{e}_{\pm}e^{-\mathrm{i}\left(\omega_{c,s}t-k_{c,s}z\right)}\right]\mbox{, }
\end{eqnarray*}
where $\mathcal{E}_{c,s}\left(z,t\right)$ denote the respective slowly-varying
amplitudes of the control and signal fields, $\omega_{c,s}$ and $k_{c,s}$
stand for their respective frequencies and wavenumbers, while $\mathbf{e}_{\pm}\equiv(\mathbf{e}_{x}\pm\mbox{i}\mathbf{e}_{y})/\sqrt{2}$
($\mathbf{e}_{x}$ and $\mathbf{e}_{y}$ define an arbitrary basis
in the plane perpendicular to the propagation direction $\mathbf{e}_{z}$).

Following e.g. \cite{GAL07}, we model the atomic sample as a continuous
medium of uniform linear density $n_{at}$, and define the average
density matrix of the slice $\left[z,z+\Delta z\right]$ by 
\[
\mathbf{\hat{\sigma}}\left(z,t\right)\equiv\frac{1}{n_{at}\Delta z}\sum_{z\leq z_{i}\leq z+\Delta z}\hat{\sigma}_{i}\left(t\right)\mbox{. }
\]

We moreover define the density matrix elements $\sigma_{ij}=\left\langle i\left|\hat{\sigma}\right|j\right\rangle $
where $i,\,j$ refer to the atomic levels and can take the values
$-1,\,1$ or $e$ (see Fig. \ref{Atomic struture}). We introduce
the slowly-varying coherences $\tilde{\sigma}_{e1},\,\tilde{\sigma}_{e-1}$
and $\tilde{\sigma}_{-11}$ defined by:
\begin{align*}
\tilde{\sigma}_{e1} & =e^{\textrm{\mbox{\mbox{i}}}\omega_{s}\left(t-\frac{z}{c}\right)}\sigma_{e1},\\
\tilde{\sigma}_{e-1} & \equiv e^{\textrm{\mbox{\mbox{i}}}\omega_{c}\left(t-\frac{z}{c}\right)}\sigma_{e-1},\\
\tilde{\sigma}_{-11} & \equiv e^{\textrm{\mbox{\mbox{\mbox{\mbox{i}}}}}\Delta_{R}\left(t-\frac{z}{c}\right)}\sigma_{-11},
\end{align*}
and write the Bloch equations in the rotating wave approximation,
for the class of velocity which is at the center of the Doppler profile:\begin{widetext}
\begin{eqnarray}
\partial_{t}\sigma_{-1-1} & = & \frac{\gamma_{t}}{3}-\gamma_{t}\sigma_{-1-1}+\frac{\Gamma_{0}}{2}\sigma_{ee}+\mbox{i}\left(\Omega_{c}^{*}\tilde{\sigma}_{e-1}-\Omega_{c}\tilde{\sigma}_{-1e}\right),\label{Bloch1}\\
\partial_{t}\sigma_{ee} & = & -\left(\Gamma_{0}+\gamma_{t}\right)\sigma_{ee}+\mbox{i}\left(\Omega_{c}\tilde{\sigma}_{-1e}-\Omega_{c}^{*}\tilde{\sigma}_{e-1}+\Omega_{s}\tilde{\sigma}_{1e}-\Omega_{s}^{*}\tilde{\sigma}_{e1}\right),\label{Bloch2}\\
\partial_{t}\sigma_{11} & = & \frac{2\gamma_{t}}{3}-\gamma_{t}\sigma_{11}+\frac{\Gamma_{0}}{2}\sigma_{ee}+\mbox{i}\left(\Omega_{s}^{*}\tilde{\sigma}_{e1}-\Omega_{s}\tilde{\sigma}_{1e}\right),\label{Bloch3}\\
\partial_{t}\tilde{\sigma}_{-11} & = & -\left(\gamma_{-11}-\mbox{i}\Delta_{R}\right)\tilde{\sigma}_{-11}+\mbox{i}\left(\Omega_{c}^{*}\tilde{\sigma}_{e1}-\Omega_{s}\tilde{\sigma}_{-1e}\right),\label{Bloch4}\\
\partial_{t}\tilde{\sigma}_{-1e} & = & -\left(\gamma_{-1e}+\mbox{i}\Delta_{c}\right)\tilde{\sigma}_{-1e}+\mbox{i}\left[\Omega_{c}^{*}\left(\sigma_{ee}-\sigma_{-1-1}\right)-\Omega_{s}^{*}\tilde{\sigma}_{-11}\right],\label{Bloch5}\\
\partial_{t}\tilde{\sigma}_{1e} & = & -\left(\gamma_{1e}+\mbox{i}\Delta_{s}\right)\tilde{\sigma}_{1e}+\mbox{i}\left\{ \Omega_{s}^{*}\left(\sigma_{ee}-\sigma_{11}\right)-\Omega_{c}^{*}\tilde{\sigma}_{1-1}\right\} .\label{Bloch6}
\end{eqnarray}
\end{widetext}

Here, $\Gamma_{0}$ is the population decay rate of the state $\left|e\right\rangle $,
and the Rabi frequencies $\Omega_{c,s}$ are defined by \textbf{
\begin{eqnarray*}
\hbar\Omega_{c,s} & \equiv & \frac{1}{2}d_{c,s}\mathcal{E}_{c,s}\left(z,t\right)\mbox{, }
\end{eqnarray*}
}where $d_{c,s}\equiv\left\langle e\left|\hat{\mathbf{d}}.\mathbf{e}_{\pm}\right|\mp1\right\rangle $
are the relevant matrix elements of the dipole operator $\hat{\boldsymbol{d}}$.

To take into account all the atoms that are distributed in different
velocity classes over the Doppler linewidth, we developed a simple
model, in which the optical coherence decay rates $\gamma_{1e}=\gamma_{-1e}$
are replaced by the effective Doppler width $\Gamma_{D}$. This gives
satisfactory results thanks to the redistribution of the pumping by
velocity changing collisions \cite{GGD08,FVA06}. All our simulations
were performed using this purely homogeneous broadening model. Consequently,
we call $\Delta_{c,s}$ the optical detuning, implicitly defined with
respect to the center of the Doppler profile.

To ensure that the full population remains constant, the discharge-assisted
ground-state feeding rate $\Lambda$ has been set to $\gamma_{t}$,
the transit rate of the atoms through the laser beam. Moreover, while
the state $\left|-1\right\rangle $ is fed with the rate $\frac{\Lambda}{3}=\frac{\gamma_{t}}{3}$,
the state $\left|1\right\rangle $ is effectively fed with the rate
$\approx\frac{2\Lambda}{3}=\frac{2\gamma_{t}}{3}$. As can be checked
by considering the full 6-level system including not only the $\Lambda$
system of interest but also the states $\left|2^{3}S_{1},m_{J}=0\right\rangle ,\left|2^{3}P_{1},m_{J}=\pm1\right\rangle $,
the state $\left|1\right\rangle $ is indeed directly fed by the rf
discharge with the rate $\frac{\Lambda}{3}$, but also indirectly
via the state $\left|2^{3}S_{1},m_{J}=0\right\rangle $ whose population
is (almost) immediately transferred to $\left|1\right\rangle $ through
optical pumping.

Finally, in the medium, the fields propagate according to the Helmholtz
equation, written in the slowly-varying envelope approximation

\begin{equation}
\left(\frac{\partial}{\partial z}+\frac{1}{c}\frac{\partial}{\partial t}\right)\Omega_{c,s}\left(z,t\right)=\mathrm{i}\eta_{c,s}\tilde{\sigma}_{e\mp1}\left(z,t\right),\label{Propagation}
\end{equation}
where $\eta_{c,s}\equiv(n_{at}k_{c,s}|d_{c,s}|^{2})/(2\hbar\epsilon_{0})$.

The set of Maxwell-Bloch equations Eqs (\ref{Bloch1}-\ref{Propagation})
was numerically solved in Matlab using the Lax discretization method
\cite{NumRec}. The medium was split into $100$ spatial steps of
$0.6\mbox{ mm}$ while the whole storage/retrieval sequence was split
into $6\times10^{6}$ timesteps of $2$ ps. We present and discuss
our numerical results in the following section.

\section{Results and discussion \label{SecIII}}

\begin{figure}
\begin{centering}
\includegraphics[width=8cm]{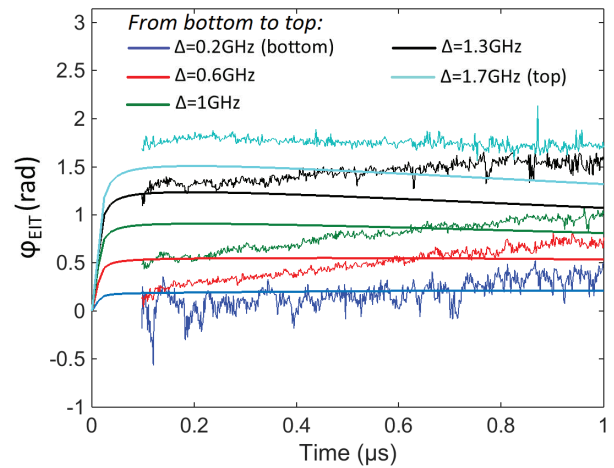}
\par\end{centering}

\protect\caption{\textcolor{black}{Temporal evolution of the extra phase shift $\varphi_{EIT}(t)$.
Experimental data, recorded from bottom to top: for $\Delta=0.2$
GHz (blue, bottom), $0.6$ GHz (red), $1$ GHz (green), $1.3$ GHz
(black), $1.7$ GHz (cyan, top). Corresponding simulations are shown
as continuous lines. The time origin corresponds to the begining of
the retrieval, when the coupling is switched on again. The storage
time is set at }\textbf{\textcolor{black}{$T=0.6\,\mu$}}\textcolor{black}{s}\textbf{\textcolor{black}{.}}}

\label{Results}
\end{figure}

All the experimental results and simulations are performed at two-photon
resonance, which means that the coupling and signal optical detunings
are equal. Fig. \ref{Results} shows experimental records for the
time-dependent extra phase shift $\varphi_{EIT}(t)$, achieved with
different values of the optical detunings $\Delta=\Delta_{c}=\Delta_{s}$,
between $0$ and $2$ GHz. The detunings are set here on the positive
side, where the nearest state ($^{3}P_{0}$) is nearly $30$ GHz away.
Each curve is obtained after averaging over 15 sets of data, recorded
at different times, for different positions of the homodyne detection
piezo-actuator. The traces recorded on the oscilloscope present some
spurious oscillations at a period of about $90$ ns. This noise is
generated by the acousto-optic modulators and could be removed by
numerically filtering the spurious frequencies during the data processing.
In Fig. \ref{Results}, the time origin corresponds to the begining
of the retrieval, when the coupling beam is turned on again. At that
time, the probe intensity starts increasing to form the retrieved
pulse. We only plot the evolution of the extra phase shift $\varphi_{EIT}(t)$
when the signal intensity is high enough, typically from roughly $100$
ns to $1\,\mu\mbox{s}$ after the start of the retrieval. One can
see that $\varphi_{EIT}(t)$ is not constant over the retrieval, and
its magnitude increases with the optical detuning $\Delta$.

The experimental plots are compared with numerical simulations of
the full Maxwell-Bloch equations derived as explained in the previous
section. These simulations are in good agreement with experimental
results: both present the same general shape for $\varphi_{EIT}\left(t\right)$,
and the same qualitative behaviour with the optical detuning $\Delta$.
One possible source for the observed discrepancies is our oversimplified
treatment of the velocity distribution in the Doppler profile. Here,
we indeed assumed that velocity changing collisions are efficient
enough to instantaneously and perfectly redistribute atoms pumped
in the probed level $\left|1\right\rangle $ over the effective Doppler
profile so that all these atoms contribute coherently to the storage
process as if the broadening were homogeneous. Although this approximation
is commonly used (see \cite{GGD08,FVA06}), it might be severely questioned
here, especially in optically detuned conditions that change the thermal
equilibrium. In particular, the absorption of the coupling beam measured
experimentally could not be well reproduced by the simulations. This
should also have an effect on the storage efficiency and on the temporal
shape of the phase. Note that in the simulation program, in order
to ``minimize'' this problem, we use an averaged coupling intensity
over the length of the cell as the input parameter, instead of the
real coupling intensity measured at the entrance of the cell. Note
also that we have checked our numerical results agree with the analytic
approximate solutions presented in \cite{GAL07}. 

\begin{figure}
\begin{centering}
\includegraphics[width=8cm]{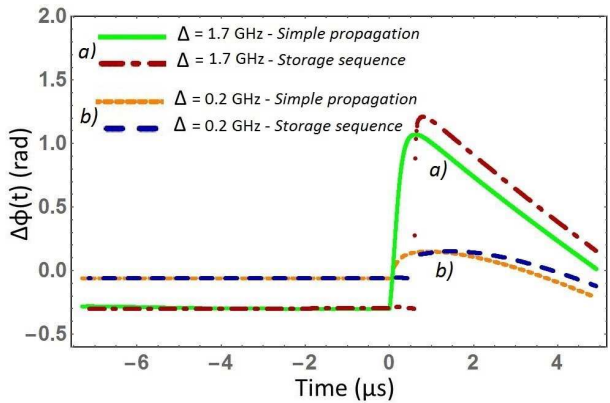}
\par\end{centering}

\protect\caption{\textcolor{black}{Temporal evolution of the relative phase $\Delta\phi(t)$
during the direct propagation of a weak signal pulse through the medium
under EIT conditions (``Simple propagation'') and during the storage
and retrieval of the same pulse (``Storage sequence''), for two
different optical detunings: a) $1.7$ GHz and b) $0.2$ GHz. The
time origin is arbitrarily chosen as the starting of the storage period. }}

\label{Susceptibility}
\end{figure}

\begin{figure}
\begin{centering}
\includegraphics[width=8cm]{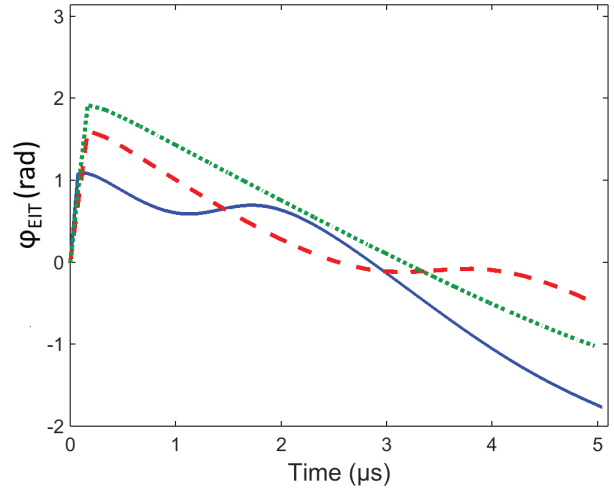}
\par\end{centering}

\protect\caption{\textcolor{black}{Temporal evolution of the phase shift $\varphi_{EIT}(t)$,
for different optical detunings in the Raman regime: $\Delta=10$
GHz (blue, full line), $\Delta=15$ GHz (red, dashed line), $\Delta=20$
GHz (green, dotted line). These curves are obtained with the full
simulation of Maxwell-Bloch set of equations, for ten times more atoms
than in our experimental case, and for a coupling power of $200$
mW. The time origin is chosen as the starting of the retrieval period. }}

\label{Raman}
\end{figure}

To understand better the physical origin of the extra phase-shift
$\varphi_{EIT}(t)$, we compared the time-dependent relative phase
$\Delta\phi\left(t\right)$ between the signal and coupling beams
at the exit of the cell, obtained: \emph{i}) during the storage and
retrieval of a weak signal pulse ($\Delta\phi(t)=\Delta\phi^{(l)}(t)$
before $t=0$ and $\Delta\phi(t)=\Delta\phi^{(r)}(t)$ after the storage
time $T=0.6\,\mu\mbox{s})$, and \emph{ii}) during the direct EIT-propagation
of the same weak pulse in the medium, while the coupling amplitude
remains constant. To compute $\Delta\phi(t)$ in case \emph{i}), we
used the full simulation of Maxwell-Bloch set of equations, whereas
in case \emph{ii}) we simply propagated each spectral component $\omega$
of the incoming pulse with the corresponding susceptibility $\chi\left(\omega\right)$.
Fig. \ref{Susceptibility} simultaneously displays the results we
obtained in both cases, for two different values of the optical detuning
$\Delta=0.2,\,1.7\mbox{ GHz}$. The shape and order of magnitude for
$\Delta\phi\left(t\right)$ are clearly the same in cases \emph{i},
\emph{ii}): the main effect of the storage is to introduce a delay
corresponding to the storage time $T$ . Here we chose $T=0.6\,\mu\mbox{s},$
but we checked both experimentally and theoretically that this phase
shift does not depend on this storage time. This suggests that the
observed extra phase shift $\varphi_{EIT}$ is essentially due to
the propagation under EIT conditions. Indeed, the stored part presents
a sharp decrease associated with many frequency components, and is
thus very sensitive to dispersive effects. This problem should be
taken into account for high speed information applications, like experiments
performed in the Raman configuration \cite{RNL10,RML11,RNJ12}. We
have thus performed simulations in the far detuned regime as presented
on Fig. \ref{Raman}. The simulation results shown here are plotted
for three different optical detunings $\Delta=10$ GHz, $\Delta=15$
GHz, $\Delta=20$ GHz, much higher than the $\Gamma_{D}\approx0.8\,$GHz
Doppler broadening. They were obtained for a number of atoms $n_{at}$
which is $10$ times higher than in our experimental case, and for
a coupling power of $200$ mW. These simulation results demonstrate
a similar effect on the retrieved signal pulse phase, even slightly
stronger than in our experimental conditions. We have also checked
that our numerical results in the Raman configuration agree with
the analytic approximate solutions presented in \cite{GAL07}.

\section{Conclusion\label{SecIV}}

In this paper, we have experimentally investigated a time-dependent
extra phase shift that appears in a storage-retrieval experiment,
performed in a room temperature atomic cell, in optically detuned
conditions. This phase shift varies with time and does not depend
on the storage time. We have provided numerical simulations which
qualitatively agree with the experimental results: in particular,
it appears that the magnitude of the relative phase depends on the
optical detuning, while its temporal shape is mainly given by the
spectrum of the incoming pulse. We explain the existence of this extra-phase
by propagation effects that can be understood by a simple propagation
model under EIT conditions with optically detuned beams. Discrepancies
may be due to an approximate treatment of Doppler broadening in the
cell. 

The results presented here might be of importance, particularly for
light storage experiments performed in the far-detuned Raman regime,
as reported in \cite{RML11}. How these results translate into the
regime of quantum light is an intriguing feature that we intend to
address in a future work. 
\begin{acknowledgments}
The work of M.-A.M. is supported by the D\'{e}l\'{e}gation Generale à l'Armement
(DGA), France and the work of J. Lugani by an Indo-French CEFIPRA
funding. We also thank the labex PALM and the R\'{e}gion Ile de France
DIM NANOK for funding.\end{acknowledgments}

\end{document}